\documentclass[conference,a4paper]{IEEEtran}
\IEEEoverridecommandlockouts
\usepackage{amsmath,amssymb,amsfonts}
\usepackage{algorithmic}
\usepackage{graphicx}
\usepackage{textcomp}
\usepackage{xcolor}
\usepackage[numbers]{natbib}
\usepackage{hyperref}
\usepackage{array}
\usepackage{listings}

\def\BibTeX{{\rm B\kern-.05em{\sc i\kern-.025em b}\kern-.08em
    T\kern-.1667em\lower.7ex\hbox{E}\kern-.125emX}}

\lstdefinelanguage{JavaScript}{
  morekeywords=[1]{break, continue, delete, else, for, function, if, in,
    new, return, this, typeof, var, void, while, with},
  morekeywords=[2]{false, null, true, boolean, number, undefined,
    Array, Boolean, Date, Math, Number, String, Object},
  morekeywords=[3]{eval, parseInt, parseFloat, escape, unescape},
  sensitive,
  morecomment=[s]{/*}{*/},
  morecomment=[l]//,
  morecomment=[s]{/**}{*/}, 
  morestring=[b]',
  morestring=[b]"
}[keywords, comments, strings]
\lstdefinelanguage[ECMAScript2015]{JavaScript}[]{JavaScript}{
  morekeywords=[1]{await, async, case, catch, class, const, default, do,
    enum, export, extends, finally, from, implements, import, instanceof,
    let, static, super, switch, throw, try},
  morestring=[b]` 
}
\lstalias[]{ES6}[ECMAScript2015]{JavaScript}
\definecolor{mediumgray}{rgb}{0.3, 0.4, 0.4}
\definecolor{mediumblue}{rgb}{0.0, 0.0, 0.8}
\definecolor{forestgreen}{rgb}{0.13, 0.55, 0.13}
\definecolor{darkviolet}{rgb}{0.58, 0.0, 0.83}
\definecolor{royalblue}{rgb}{0.25, 0.41, 0.88}
\definecolor{crimson}{rgb}{0.86, 0.8, 0.24}
\lstdefinestyle{JSES6Base}{
  backgroundcolor=\color{white},
  basicstyle=\ttfamily,
  breakatwhitespace=false,
  breaklines=false,
  captionpos=b,
  columns=fullflexible,
  commentstyle=\color{mediumgray}\upshape,
  emph={},
  emphstyle=\color{crimson},
  extendedchars=true,  
  fontadjust=true,
  frame=single,
  identifierstyle=\color{black},
  keepspaces=true,
  keywordstyle=\color{mediumblue},
  keywordstyle={[2]\color{darkviolet}},
  keywordstyle={[3]\color{royalblue}},
  numbers=left,
  numbersep=5pt,
  numberstyle=\tiny\color{black},
  rulecolor=\color{black},
  showlines=true,
  showspaces=false,
  showstringspaces=false,
  showtabs=false,
  stringstyle=\color{forestgreen},
  tabsize=2,
  title=\lstname,
  upquote=true  
}
\lstdefinestyle{JavaScript}{
  language=JavaScript,
  style=JSES6Base
}
\lstdefinestyle{ES6}{
  language=ES6,
  style=JSES6Base
}

\begin{document}

\newcommand{\cooe}{CO\textsubscript{2}e }
\newcommand\rev[1]{#1}

\title{The Environmental Costs of Surveillance Capitalism: A Case Study of Social Media Platforms}

\author{\IEEEauthorblockN{Nils Bonfils}
\IEEEauthorblockA{\textit{University of Toronto}\\
Toronto, Canada \\
\href{mailto:nils.bonfils@mail.utoronto.ca}{nils.bonfils@mail.utoronto.ca}}
\and
\IEEEauthorblockN{Christoph Becker}
\IEEEauthorblockA{\textit{University of Toronto}\\
Toronto, Canada \\
\href{mailto:christoph.becker@utoronto.ca}{christoph.becker@utoronto.ca}}
}

\maketitle

\begin{abstract}
The business model of surveillance capitalism, premised on the extraction of behavioral data and its predictive potential for profit, relies on extensive material infrastructure. \rev{Such profit is typically driven by practices such as telemetry, user tracking, data analytics, secondary data uses, increased user engagement,} and AI model training, as well as large-scale data storage systems that retain personal information for sale or reuse.


This paper is motivated by the question: how much of the rising carbon impact of ICT can be attributed to this material infrastructure? Such an inquiry provides a foundation for quantifying the environmental costs of surveillance capitalism by proposing a conceptual framework and research direction that link processes of surveillance with their underlying material realities.
To demonstrate the applicability of this framework, we examine the proportion of \rev{network traffic caused by surveillance capitalism processes} through a comparative case study of a corporate social media platform, X/formerly Twitter, and a decentralized, non-commercial alternative, Mastodon. 

Our findings highlight the existence of \textit{corporate overhead}: excess resource consumption driven by corporate social media practices, which \rev{is used as an initial proxy for the activities of surveillance capitalism. Our findings further demonstrate how the corporate overhead of X can be used to establish a lower bound in \cooe emissions attributable to for-profit activities that do not contribute to the user experience.}
\end{abstract}


\section{Introduction}

Digital platforms, such as social media, have transformed global information dissemination and have become a defining technology of the age of information. Since Facebook's launch in the late 2000s, platforms like YouTube, Instagram, and TikTok have achieved near-ubiquitous reach. For-profit companies have established the dominant business model for social media platforms: the commodification of their users through targeted advertising and data extraction \rev{\cite{couldryDataColonialismRethinking2019b}.} 

\rev{These incentives have given rise to a range of pervasive practices in digital life, including infrastructure that tracks user behavior, algorithmic feeds designed to maximize engagement, and the centralization of power in corporations. For the purpose of this paper, we adopt the concept of surveillance capitalism \cite{zuboffAgeSurveillanceCapitalism2019}, which extends beyond surveillance and user tracking and includes activities such as user engagement optimization through rich media and advertisements, to describe the processes and practices underpinning the dominant business model.}

\rev{The social harms of surveillance capitalism processes have been recognized and documented \cite{aiolfiDatadrivenDigitalAdvertising2021,hamRolePersuasionKnowledge2016,wuSlowViolenceSurveillance2023b}.} However, these processes also carry significant environmental implications due to the material infrastructure required to sustain them. The rapid expansion of information and communication technology (ICT), compounded with the computational demands of surveillance capitalism, run the risk of becoming major contributors to ICT’s growing environmental footprint. For instance, the processing power required for behavioral predictions, servers that store vast amounts of user data, and networks that transmit algorithmically targeted content all incur substantial material and environmental costs. Data centers alone account for 1.5 percent of global electricity consumption \cite{EnergyAI2025}, with projections suggesting this figure could double by 2030. As the climate crisis intensifies, the environmental footprint of Corporate Social Media (CSM) platforms demands urgent scrutiny alongside their social impacts.

Alternative social media (ASM) platforms emerged as a direct response to surveillance capitalism. Beginning with Diaspora in 2010 and accelerating after major scandals such as Cambridge Analytica and Elon Musk's acquisition of Twitter, these platforms represent a re-imagining of digital communication. In contrast to the business model of CSMs, ASMs are based on federated architectures, open protocols, nonprofit governance, and community-controlled moderation \cite{gehlCaseAlternativeSocial2015,schneiderGovernableSpacesDemocratic2024}.

\rev{Because ASMs do not rely on surveillance or engagement-maximization infrastructures in order to offer functionally equivalent services, they provide a useful empirical baseline for comparison. In effect, ASMs more typically pursue digital sufficiency -- an approach that prioritizes what is necessary to achieve a goal rather than what is possible, which results in lower absolute resource consumption of ICTs. In this context, ASMs can be understood as providing a functionally sufficient baseline against which CSMs can be compared. The introduction of two novel metrics, \textit{corporate overhead} and \textit{user tracking overhead}, as proxies for the excess computational resources associated with surveillance capitalism processes, provides a quantitative basis for evaluating sufficiency-based strategies for minimizing ICT resource consumption.} 

This paper addresses a significant gap in the literature by linking the mechanisms through which CSMs generate profits, such as the capture, analysis, and prediction of user behavior, with their material costs. Our contribution is twofold. First, we propose a conceptual framework that situates the processes of data capture and commodification within the infrastructure of digital platforms. This conceptual framework also functions as a research agenda, in which its components represent distinct "pieces of the puzzle" for answering the broader question of the environmental costs of surveillance capitalism.
Second, we provide an initial investigation of this research agenda through a case study. Specifically, we examine the question of digital sufficiency in social media by comparing a CSM, X (formerly Twitter), and an ASM, Mastodon.

Ultimately, our contributions are guided by two research questions: (1) \textit{How can we quantify the environmental impact of surveillance capitalism?} (2) \textit{What is the proportion of resources used by surveillance capitalism in the use of digital platforms?}

\section{Background}

\subsection{Surveillance Capitalism and Its Material Impacts}

Zuboff's concept of surveillance capitalism provides a theoretical framework for understanding the political and economical mechanisms underlying the rapid growth and market dominance of CSMs, such as the Big Tech companies, whose annual revenues rival the GDP of several European countries \cite{bonfilsEmpiricalInquirySurveillance2025}. These mechanisms involve the extraction and analysis of \textit{digital traces} (also referred to as metadata or \textit{behavioral surplus}), the commodification of these traces through the creation of \textit{prediction products}, and the sale and use of these prediction products to deliver targeted services to the users of digital platforms \cite{zuboffAgeSurveillanceCapitalism2019}. \rev{In other words, surveillance capitalism can be understood as processes and practices that capture and leverage user data and metadata for profit in the form of increased user engagement, captured attention, and targeted advertisements.} This constitutes the foundation of CSM's business model and, in some cases, grants them sufficient political leverage to shape favorable public policy \cite{zuboffBigOtherSurveillance2015,zuboffAgeSurveillanceCapitalism2019}. While the literature has drawn attention to the problematic and unsustainable nature of the infrastructure around surveillance capitalism \cite{landwehrProblemsSurveillanceCapitalism2023b}, it has remained limited in scope. In particular, the elements presented in Zuboff's theory lack direct empirical grounding in material infrastructural realities \cite{greenleafElementsZuboffsSurveillance2019}.

Within ICT4S, big data analytics has been researched for its potential to yield insights into climate related challenges \cite{armbrusterBigDataBig2015,shahrokniBigDataGIS2014}. For our purposes, analytics corresponds to the surveillance capitalism processes of data extraction, analysis, and prediction operationalized through data gathering, data mining, and the evaluation and interpretation of the results \cite{tsaiBigDataAnalytics2015}. Furthermore, Big Tech companies like Facebook have openly shared how their analytics infrastructure is tied to advertising and user tracking activities \cite{thusooDataWarehousingAnalytics2010}. While there has been a significant effort to improve the energy efficiency of big data analytics systems \cite{arulEnergyefficientDataEngineering,dayarathnaEnergyConsumptionAnalysis2017}, these systems nevertheless have a substantial measurable footprint. The literature on big data analytics provides concrete elements, such as specific software systems and technologies, through which ICT infrastructure underpinning surveillance capitalism can be understood \cite{grossmanWhatAnalyticInfrastructure2009,arfatBigDataTools2020}.

\subsection{Digital Sufficiency}\label{sec:sufficiency}

ICTs are often framed as technologies that support and contribute to human flourishing while also improving the environmental crisis; however, they are increasingly understood as placing additional pressure on environmental systems rather than alleviating it. In an attempt to enable ICTs to play a more beneficial role in supporting life and ecosystems on our planet, Santarius et al. introduced the concept of digital sufficiency and its four dimensions\rev{: hardware sufficiency, software sufficiency, user sufficiency, and economic sufficiency} \cite{santariusDigitalSufficiencyConceptual2023}. Sufficiency is defined as any strategy that aims to reduce resource and energy use by lowering levels of production and consumption. Recent ICT4S publications have examined digital sufficiency across three of the four dimensions of digital sufficiency. Szalkowski \& Windekilde have explored both \textit{hardware sufficiency} and \textit{economic sufficiency} through a simulation of a policy scenario to reduce e-waste, extend hardware lifespan, and encourage electronics repair and reuse \cite{szalkowskiICTSufficiencyNecessary2024}. Gatt et al. focused on \textit{user sufficiency} by modeling the efforts from users who were environmentally aware to minimize energy consumption in critical periods by when renewable energy was not available \cite{gattDigitalSufficiencyBehaviors2024}.

Beyond ICT4S, the field of sustainable HCI has also identified one of the major unsustainable aspects of ICT use and design as the \textit{cornucopian paradigm}. Without explicitly invoking the term, Widdick and Pargman advance an argument for sufficiency by advocating for the moderation of internet use \cite{widdicksBreakingCornucopianParadigm2019a}. \rev{Moderating internet use is a form of user sufficiency that may directly reduce the volume of data extracted and the level of user engagement promoted by surveillance capitalism processes.} In their work, Preist et al. have initially identified the cornucopian paradigm and similarly advance a sufficiency-oriented argument by characterizing unwelcome distraction in the form of video advertisements, \rev{(an important element of surveillance captialism processes)}, as a form of digital waste \cite{preistUnderstandingMitigatingEffects2016}. \rev{Digital sufficiency provides another important framework for understanding the sustainability implications of surveillance capitalism.} This is particularly relevant given that Santarius et al. also identify a part of ICT's environmental burden as the exponential increase in data storage, processing, and transmission driven by big data analytics \cite{santariusDigitalSufficiencyConceptual2023}.

\subsection{Measuring the Environmental Impact of Digital Platforms}

In order to support debates on sufficiency in big data analytics, it is essential to quantify the share of energy and resources used by analytics in digital platforms. Quantifying and attributing resource consumption and environmental impact is challenging. However, existing environmental assessments have quantified the carbon footprint of digital content consumption \cite{istrateEnvironmentalSustainabilityDigital2024a}, of specific corporate social media platforms \cite{batmunkhCarbonFootprintMost2022b}, and even of tracking cookies.\footnote{\url{https://carbolytics.org/report.html}} Recent ICT4S literature has also proposed a method to measure the client-side energy consumption caused by web analytics scripts present on websites \cite{puhtilaEffectAnalyticsTools2024}. To date, no studies have examined how equivalent and sufficient alternative digital platforms compare to their corporate counterparts or how they might reduce computational resource use and energy demands.


\section{Mapping the Environmental Costs of Surveillance Capitalism}
\label{sec:framework}
Quantifying the environmental impacts of surveillance capitalism is a complex task. In part, this is due to the imperative of surveillance capitalists, such as CSMs, to maintain a knowledge asymmetry. It is also because assessing environmental impacts is inherently a challenging process. It involves multiple orders of interrelated effects impacting complex and dynamic ecological systems that are still not well-understood to this day. In order to lay the foundation for such an undertaking, we propose a framework that maps out the different conceptual elements to the material impacts of the infrastructure underlying digital platforms.

Because the investigation of the environmental costs of surveillance capitalism is a question of digital sufficiency, it is important to consider the share of computational resources that surveillance capitalism consumes in digital platforms. Accordingly, we define three broad categories of computational resource usage from a functional point of view: \textit{core \& navigation features}, \textit{user generated data \& content}, and \textit{surveillance \& tracking}.

\begin{figure}[h]
  \centering
  \includegraphics[width=0.85\linewidth]{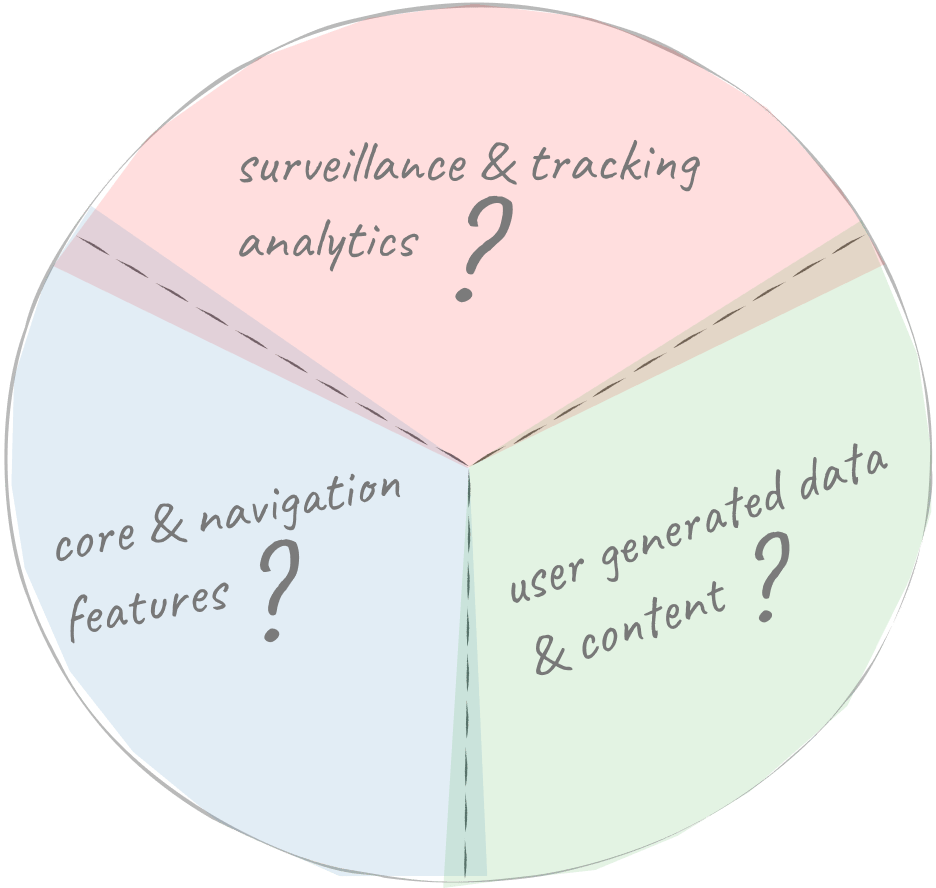}
  \caption{Each of the three categories represent the share of resource usage by functionalities related to that category. Our central question is simply: How large is each part?}
  \label{fig:resourceproportion}
\end{figure}

The overarching questions then are: 1) what is the share of environmental costs caused by the sufficient functionalities of the platform? and 2) what is the share of of environmental costs caused by surveillance capitalism processes (see fig. \ref{fig:resourceproportion})? 

\begin{figure*}[h]
  \centering
  \includegraphics[width=\linewidth]{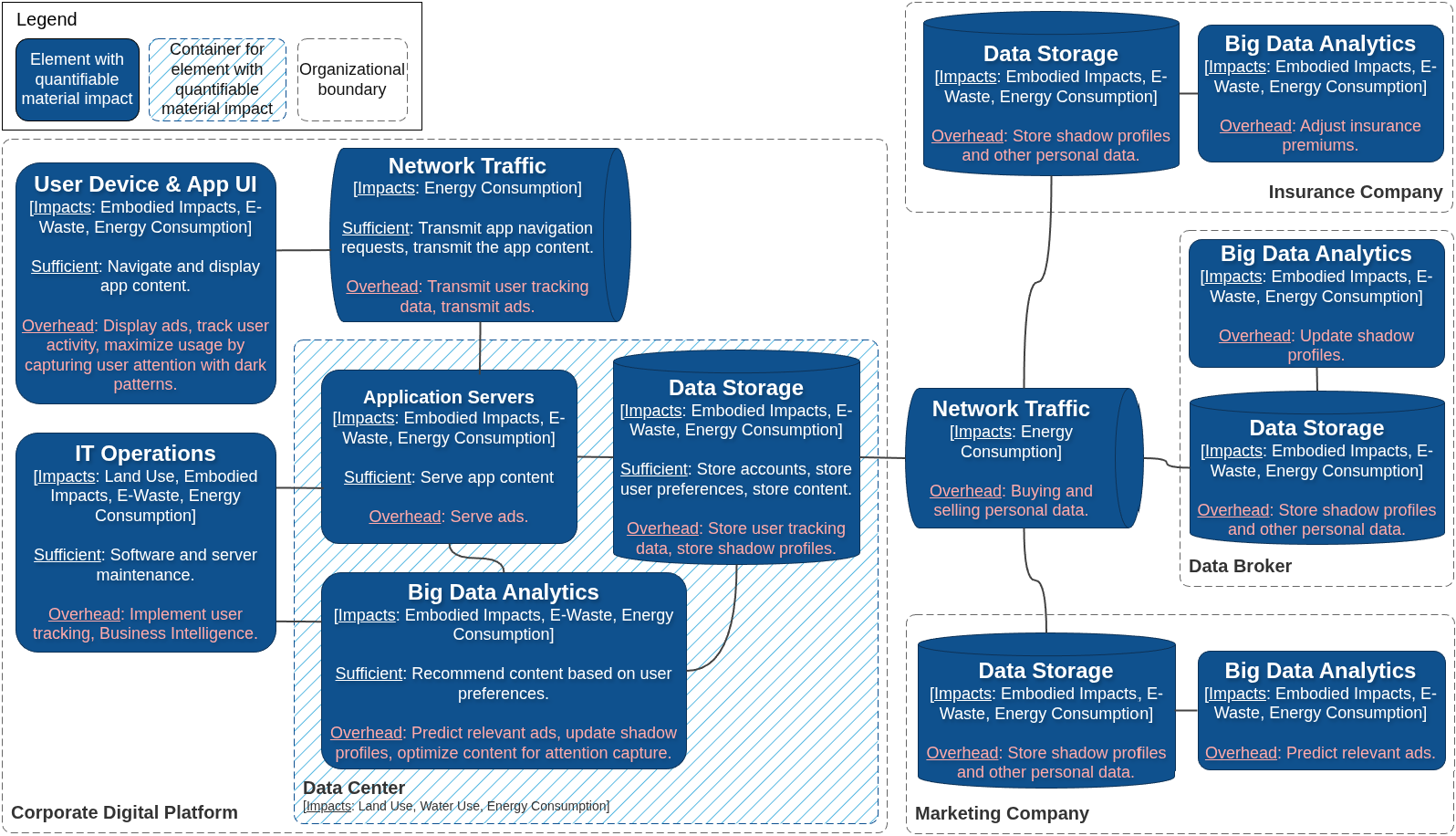}
  \caption{The conceptual framework that maps the constituent elements of a digital platform engaged in surveillance capitalism to their measurable material impacts. The text in the plain blue boxes contains the following textual elements (from top to bottom): a title, a list of associated material impacts in brackets, a list of sufficient functionalities, and a list of typically undesirable functionalities that are useful to surveillance capitalism in red. The lines have no particular meaning besides indicating a relationship or connection between two elements.}
  \label{fig:framework}
\end{figure*}

The primary goal of our proposed framework is to establish a conceptual foundation that links the mechanisms of data capture, processing, and commodification that are central to surveillance capitalism with their material realities. The diagram in figure \ref{fig:framework} represents the different conceptual elements of a digital platform's infrastructure. The diagram style is inspired by the C4 software architecture notation.\footnote{\url{https://c4model.com/}} The main elements of this framework, the plain blue boxes, are elements that can be associated with a quantifiable material impact. The blue hatched boxes are a helpful conceptual grouping that can also be associated with material impacts. Lastly, the dashed boxes represent organizational boundaries for environmental cost accountability. While the diagram itself is reminiscent of software architecture diagrams, it is important to keep in mind that it is intended here to serve as a conceptual map. 

Each element is attributed both \textit{sufficient} and \textit{overhead} functionalities in line with the digital sufficiency argument presented in section \ref{sec:sufficiency}. For instance, the user device \& application UI has a sufficient functionality of allowing users to navigate the app and explore its content. When accounting for surveillance capitalist processes this element would exhibit functionalities of displaying advertisements, tracking user activity, and leveraging manipulative design patterns \cite{grayDarkPatternsSide2018} to maximize user attention and time spent using the platform. Following a sufficiency-based approach to limit the increasing environmental impacts of ICTs, the elements exhibiting both sufficient and overhead functionalities should not be entirely removed but modified to keep the sufficient functionalities. Conversely, elements exhibiting only overhead functionality, such as the network traffic due to the purchase and sale of personal and behavioral data, along with the infrastructure to support the storage and processing of this data by third parties, should be phased out.

Our proposed framework is also meant to serve as a research agenda. Each element of the diagram represents a "piece of the puzzle" to solve in order to quantify the environmental cost of surveillance capitalism. As most elements contain both sufficient and overhead functionalities, one of the main challenges of quantifying the environmental overhead of these elements will be to determine the share of energy and resources attributed to the undesirable functionalities. It should be noted that our framework focuses on the first level of the LES model: Life-cycle impacts \cite{hiltyICTSustainabilityEmerging2015}. This focus means that the framework could be extended to include enabling and structural impacts, offering a more comprehensive assessment of surveillance capitalism's environmental costs. To illustrate its applicability, we follow with a case study evaluating the environmental costs associated with X/Twitter's network traffic between its application UI and its application servers.

\section{Case Study}

Answering the difficult question of measuring the environmental costs of surveillance capitalism requires an incremental approach. From our conceptual framework (see fig. \ref{fig:framework}), the elements that are most visible and accessible to the public are the user device \& application UI and the network traffic that is transmitted to the application servers. While it should be possible to investigate both elements at the same time, for the sake of simplicity, we focused on network traffic. 

\begin{figure}[h]
  \centering
  \includegraphics[width=\linewidth, height=5cm]{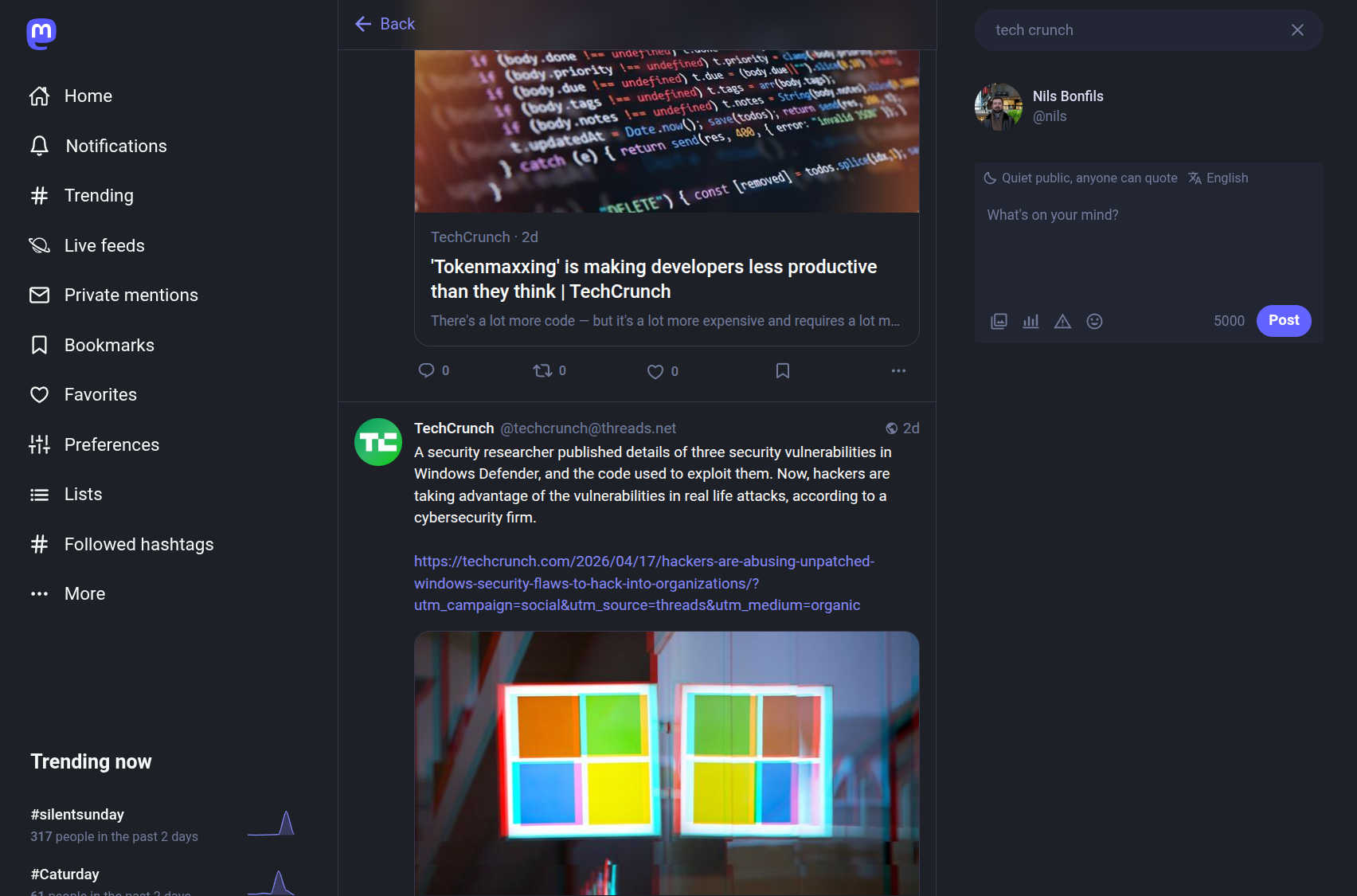}

  \vspace{2px}
  \includegraphics[width=\linewidth, height=5cm]{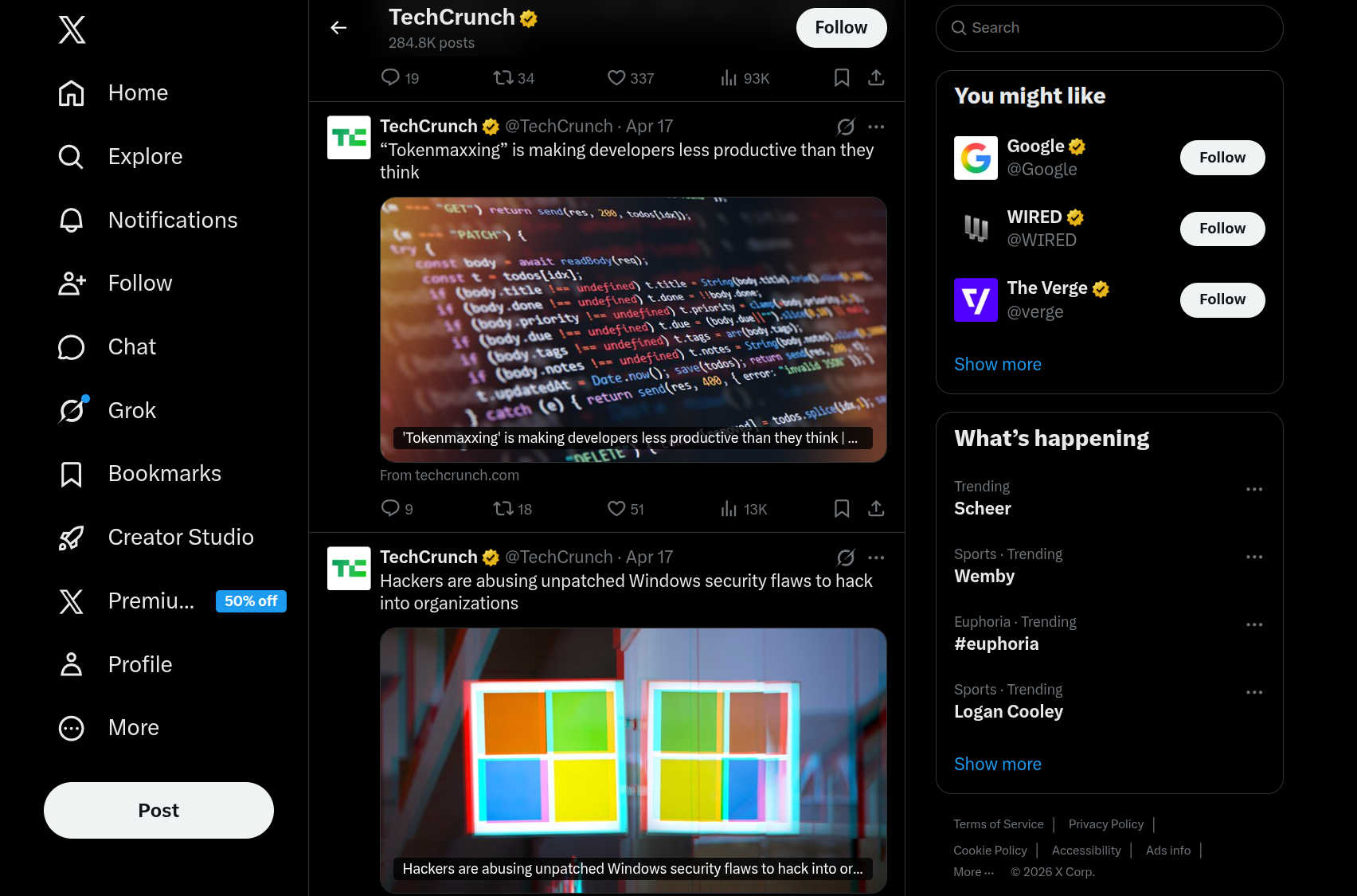}
  \caption{\rev{Mastodon (top) and X/Twitter (bottom) are visually similar, employing the well-known \textit{3-columns layout}.}}
  \label{fig:platformsux}
\end{figure}

To do so, we conducted a preliminary comparative analysis of two digital platforms: the CSM, X, and its ASM counterpart, Mastodon. \rev{Both of these platforms provide equivalent microblogging capabilities, have the same features (see table \ref{tab:platformsfeatures}), and feel very similar down to their UX design (see fig. \ref{fig:platformsux})}. However, X operates under corporate imperatives of profit maximization and growth, enacted through user data extraction, targeted advertising, and algorithmic feeds designed to increase engagement. Mastodon, on the other hand, is a federated social media based on ActivityPub and is available as free and open-source software (FOSS).

\begin{table}[htbp]
    \centering
    \caption{\rev{Non-exhaustive list of features for both platforms showing their functional equivalence as microblogging platforms.}}
    \begin{tabular}{ | c | c | c | }
        \hline
        \textbf{Feature} & \textbf{X} & \textbf{Mastodon} \\
        \hline\hline
        Post & Tweet & Toot \\
        \hline
        Re-Post & Retweet & Boost \\
        \hline
        Like & Like & Favorite \\
        \hline
        Editing & Supported (premium) & Supported \\
        \hline
        Media & Images, Videos, Audio & Images, Videos, Audio \\
        \hline
        Feed & Algorithmic & Chronological \\
        \hline
        Monetization & Ads, Subscriptions & None, Donations \\
        \hline
    \end{tabular}
    \label{tab:platformsfeatures}
\end{table}

The fundamental distinction between both platforms lies in their underlying economic incentives. In other words, if we consider the resource used by Mastodon and X as $res_M$ and $res_X$ respectively, and the corporate overhead in resource use due to practices such as user tracking or maximizing user attention as $res_{CO}$, we can express our assumption as follows:

\begin{equation}
    res_X = res_M + res_{CO}
\end{equation}

Investigating the environmental impact of surveillance capitalism through this lens thus becomes measuring the difference between both platforms.


\subsection{User Journeys}

User journeys, adapted from the field of UX design, are the sequences of technically defined steps that a user undertakes to complete specific actions. By instrumenting web browsers to execute user actions in an automated and controlled fashion, user journeys enable explicit, reproducible, and rigorous measurements from a realistic client-side perspective. They also serve as a practical mechanism for quantifying the differences in resource usage between two digital platforms.

User journeys were initially inspired by the OpenWPM open-source web privacy measurement tool \cite{englehardtOnlineTracking1millionsite2016a}. OpenWPM is a tool that provides useful features for detecting and measuring user tracking at scale. As it focuses on speed and breadth, it is well-suited to crawl a large number of websites. While it works particularly well at scale, it requires a non-trivial setup to function and is not designed for the purpose of navigating around a single web application or reproducing actions like a user. Furthermore, the types of measurements it provides are not adapted for our case study.

\begin{figure}[h]
  \centering
  \includegraphics[width=\linewidth]{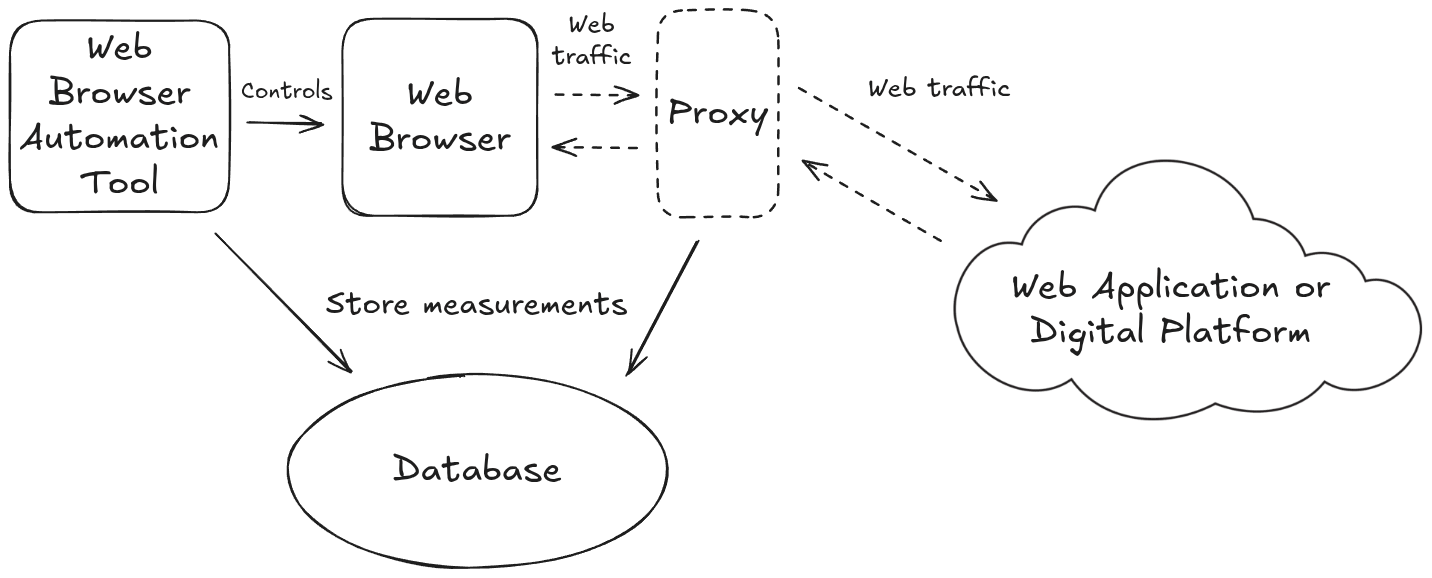}
  \caption{High level overview of the user journey measurement tool design.}
  \label{fig:userjourneytool}
\end{figure}

Building a bespoke user journey measurement tool facilitated the development of a flexible system tailored specifically for the design of user journeys and the measurement of the desired metrics. As seen in figure \ref{fig:userjourneytool}, the design of our user journey measurement tool consists of three components: a web browser automation tool or framework (also called a WebDriver), an HTTP proxy, and a database. The WebDriver controlled the web browser programmatically, typically using a scripting language such as Python or JavaScript. The WebDriver script defined the sequence of actions, including logging in, navigating by clicking on buttons or links, and scrolling on the page, among other possible actions. The proxy is used to intercept, capture, and store the network traffic which is the real representation of the user journey. Finally, the database was used to store the captured data in a structured form, enabling further analysis.

The type of data that is captured by the WebDriver script and the proxy depends on the analytical objectives of the user journey. This could vary from the content that is being displayed on screen to the timestamp of each action. 
To measure the environmental impact of surveillance capitalism, we capture metrics, such as the total duration of the user journey and all the network requests and responses with their HTTP headers (containing tracking information such as cookies) (see section \ref{sec:metrics}).

\subsection{Experimental Setup}

We measured various actions a user would undertake while using the platform. The user journeys we measured are described in table \ref{tab:userjourneys}. \rev{User journey 1 emulates a user scrolling on the main feed for approximately 5 mins. User journey 2 emulates a user writing a post and then clicking on it to open it. Finally, user journey 3 is similar to user journey 1, except that it scrolls on the feed of TechCrunch. TechCrunch is a popular account that displays equivalent content on both X\footnote{\url{https://x.com/TechCrunch}} and Mastodon.\footnote{\url{https://mastodon.social/@techcrunch@threads.net}}\textsuperscript{,}\footnote{The minor difference is that the TechCrunch feed on Mastodon contains posts with a slightly longer excerpt. It can be seen in figure \ref{fig:platformsux}.}} We decided to focus on scrolling and posting as the two main interactions. In order to obtain more stable results, each user journey was repeated \rev{100 times, representing around 75 hours spent browsing and posting on social media. The aggregated results, such as the total transfer size, were then averaged over those 100 runs.}

\begin{table}[htbp]
    \centering
    \caption{\rev{The list of user journeys.}}
    \begin{tabular}{ | c | c | p{0.6\linewidth} | }
        \hline
        \textbf{User Journey} & \textbf{Task} & \textbf{Description} \\
        \hline\hline
        \#1 & Scroll & The user scrolls for approximately 5 minutes on the main feed. \\
        \hline
        \#2 & Post & The user writes a post and clicks on the post to open it. \\
        \hline
        \#3 & Scroll & The user scrolls for approximately 5 minutes on the TechCrunch feed. \\
        \hline
    \end{tabular}
    \label{tab:userjourneys}
\end{table}

\begin{figure}[t]
  \centering
  \includegraphics[width=\linewidth]{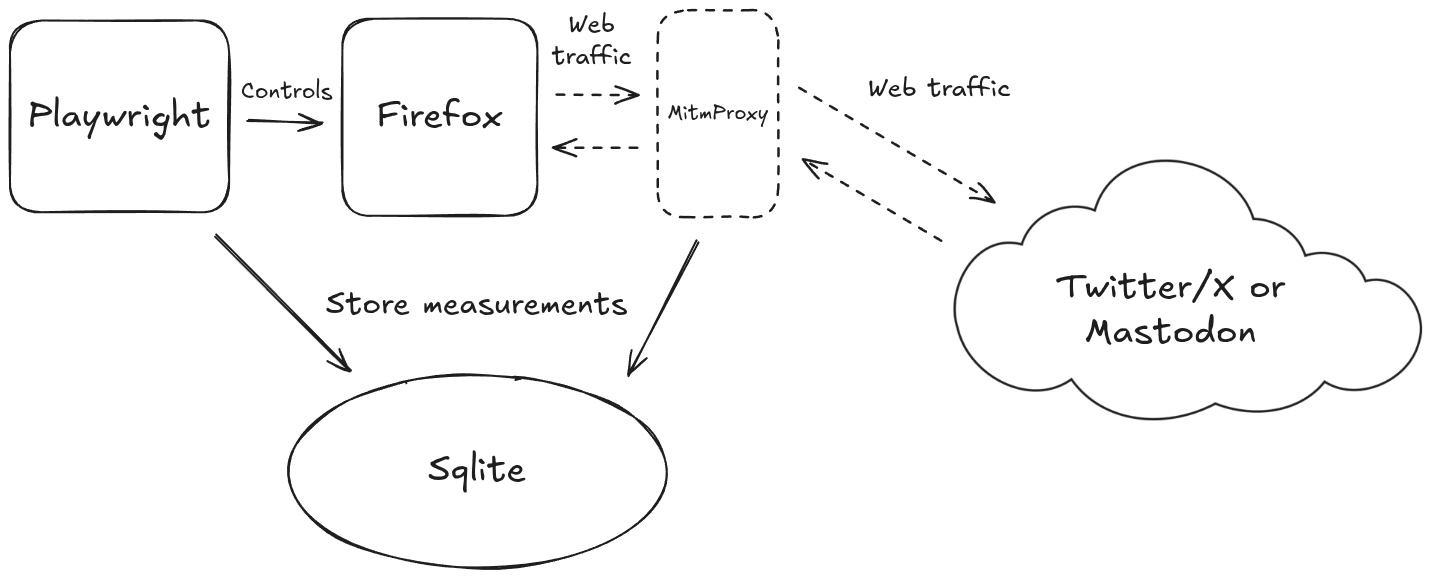}
  \caption{Concrete implementation of the user journey measurement tool used for the case study.}
  \label{fig:userjourneytoolimpl}
\end{figure}

For our case study we implemented a user journey tool using the various software depicted in figure \ref{fig:userjourneytoolimpl}. This provides a concrete example of an implementation of user journeys to measure surveillance capitalism activity on digital platforms from the bottom-up \cite{bonfilsCodeEnvironmentalCosts2026}\footnote{The code is available at \url{https://doi.org/10.5281/zenodo.20384405}}.

Playwright is a JavaScript tool that was designed to run \textit{integration tests} when developing web applications, but it can also be used as a general web browser instrumentation framework. Playwright works by enabling us to write code that takes control of a web browser (in our case Firefox) and executes actions like clicking, scrolling, and filling input forms. It allows for fairly readable scripts that act as 'recipes' for user journeys (see Listing \ref{lst:userjourney2}).

MitmProxy is an HTTPS proxy that allowed us to capture all the traffic going to and fro the web browser in which the user journey is running \cite{mitmproxy}. MitmProxy can be scripted with python, which allows us to store the captured HTTP requests and responses in the database (in our case sqlite). 

\begin{figure*}
\begin{lstlisting}[style=ES6, caption={The JavaScript code for the user journey 2 on Mastodon, making a post.}, label={lst:userjourney2}]
// Some setup code to obtain the 'page' from a web browser precedes..

await page.goto('https://mastodon.bsd.cafe');
await page.waitForTimeout(500);

const post = 'This is an automated test post that will be deleted in a few minutes.';
await page.locator('form.compose-form textarea').pressSequentially(post, {delay: 100});
await page.waitForTimeout(500);

await page.getByRole('button', { name: 'Post' }).click();
await page.waitForTimeout(500);

await page.getByText(post).click();
await page.waitForTimeout(1000);
await page.waitForLoadState('networkidle');

console.log('Closing browser');
await browser.close();
console.log('Done!');
\end{lstlisting}
\end{figure*}


\subsection{Environmental Impact Quantification}
\label{sec:envimpactquant}

Our focus in this case study is to establish a lower bound for the environmental impact of digital platforms' network traffic  attributed to surveillance capitalism activity. As noted by Billstein et al., bottom-up approaches often lead to underestimations \cite{billsteinLifeCycleAssessment2021}. Underestimates are preferable to overestimates as we are interested in establishing the presence of a non-negligible impact of surveillance capitalism activity through a lower bound. While network traffic is as an imperfect proxy for environmental impact --- given the non-linear relationship between data transfer and \cooe emissions --- it remains a useful and readily deployable metric for demonstrating that environmental consequences exist. 


In line with Istrate et al., we consider "that the electricity intensity of data centres and the core network is proportional to the load and, consequently, is expressed as the amount of electricity required \textbf{per unit of data traffic} (i.e., $kWh/GB$)." \cite{istrateEnvironmentalSustainabilityDigital2024a}. User journeys can be used to measure the network traffic generated by user interactions with a digital platform expressed in $GB$. It is then possible to derive the amount of data transferred per unit of time spent on the platform (and notably the data transferred strictly related to user tracking); \rev{we thus focus here on quantifying data volumes transferred as the key measurement.}
Multiplying this amount by the number of daily active minutes on the platform can give us an estimate of the scale of total data transferred on the platform. Using a coefficient for the energy intensity of the internet, we can derive the energy used per unit of time. Finally, we can multiply our result by the coefficient of the carbon intensity of the global electricity grid to obtain a lower bound for the carbon emission attributable to surveillance capitalism activity of our case study, X.

In order to transform the daily amount of data (in $GB$) to energy usage (in $kWh$), we use the reasonable coefficient estimated by Istrate et al. in their life cycle inventory data from the supplementary information. $0.0177\ kWh/GB$ for the core network and $0.0414\ kWh/GB$ for data centers, which gives us a final coefficient of \textbf{0.0591 kWh/GB} \cite{istrateEnvironmentalSustainabilityDigital2024a}. This coefficient is close to the one given by Schien et al. of $0.052\ kWh/GB$ \cite{schienEnergyIntensityInternet2015}. However, Schien et al.'s coefficient does not include data centers and is a decade old. As a result, the more recent estimate by Istrate et al. is to be preferred\cite{koomeyDoesNotCompute2021}. The global average of the electricity grid carbon intensity is \textbf{445 g\cooe/kWh} \cite{Electricity20252025}.

Network traffic usage metrics are derived from user journeys, they represent the network traffic from a single user. \rev{Since user journey 1 traffic is highly volatile due to the algorithmic feed of X, we used user journey 3 to estimate the traffic from browsing content as it provides a more reliable basis for comparison between X and Mastodon.} To obtain \cooe estimates that are meaningful, we need to adjust our measurements to the scale of X. There are two useful numbers in this case: X publicly disclosed \textbf{8 billion daily active minutes}\footnote{\url{https://x.com/XData/status/1769826757341802982}} and X's former CEO stated in 2023 that the platform was seeing \textbf{500 million tweets per day} \cite{perezActuallySees500M2023}.

\subsection{Metrics}\label{sec:metrics}

Our experimental setup, using our bespoke user journey measurement tool, allows for the collection of key metrics to analyze network traffic due to surveillance capitalism. Broadly speaking, our setup captures HTTP requests and responses transmitted during the user journey. The following paragraphs describe the metrics that are the most useful in our analysis.

\paragraph{URL}
URLs contain a wealth of information that is sometimes not structured or explicit. For starters, URLs are useful to distinguish the various requests during user journeys. They can also contain information about the type of content being retrieved or sent in the form of a file extension (e.g. \verb|.json| for JSON files or \verb|.jpg| for jpeg image files). Because of the way APIs are organized, URLs can also contain information about how certain services are designed, potentially leaking information about which endpoints are used for tracking user activity.

\paragraph{Cookies}
Cookies are an important mechanism of the HTTP protocol. They offer the ability to keep track of user sessions while navigating the web. Logging in and out of online services is possible because of cookies. However, that mechanism is being abused by online services with corporate interests to track user activity even when they are not logged into any services. It has even been shown that this abuse could have a non-negligible carbon footprint.\footnote{\url{https://carbolytics.org/report.html}} Cookies play a crucial role in the surveillance capitalism apparatus.

\paragraph{Total Size}
The total sizes of the HTTP requests and responses are important to keep track of how much data is being transmitted over the network.

\paragraph{Timestamp}
Keeping track of the timestamp of each HTTP request and response is essential to obtain the total duration of the user journey from the first to the last response. 

\section{Results}

Comparing X and Mastodon through user journeys provides valuable insights.
Mastodon has no corporate incentive to track its users for the purpose of profit and growth. Consequently, surveillance capitalism processes, such as user tracking and targeted advertising, are not present on the platform. This allows us to assume that Mastodon represents a \textit{sufficient baseline} for the minimum amount of data necessary for the platform to provide microblogging functionalities. It is important to underscore that the level of engineering effort invested in Mastodon is substantially lower than that of X, and therefore certain inefficiencies may exist on Mastodon that are not present on X. As a result, Mastodon likely represents an overestimate of this sufficient baseline.

\subsection{Quantifying Surveillance Capitalism Network Overheads}

We distinguish two alternative ways to examine network traffic related to surveillance capitalism activity: 1) \textit{corporate overhead} and 2) \textit{user tracking overhead}. 

\subsubsection{Corporate Overhead}
\rev{Corporate overhead is the proportion of network traffic in the CSM beyond the sufficiency baseline network traffic established by the ASM. It typically comprises user tracking and business analytics to support growth and profit, but also extends to heavier content delivery and platform features designed to increase user engagement and attention, such as richer media content and enhanced interactivity.} As noted above, the Mastodon baseline in our case likely overestimates sufficient resource usage, which implies that corporate overhead should be interpreted as an underestimate.

\begin{figure}[h]
  \centering
  \includegraphics[width=\linewidth]{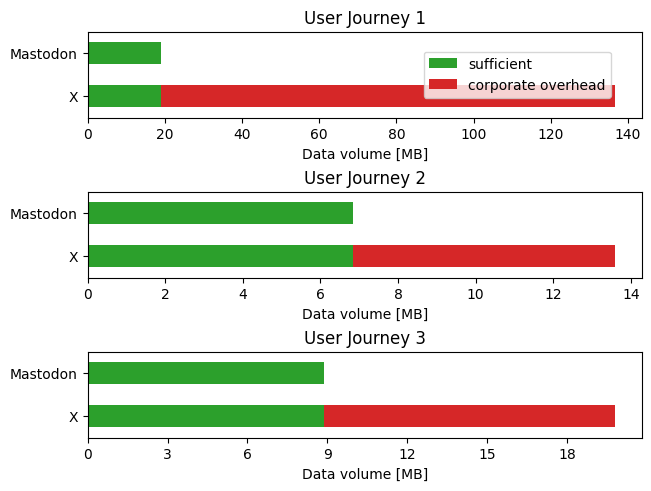}
  \caption{\rev{Bar chart highlighting the corporate overhead for user journeys 1-3.}}
  \label{fig:corporateoverhead}
\end{figure}

As can be seen in figure \ref{fig:corporateoverhead}, the corporate overhead represents \rev{86.04\% of X's data traffic when scrolling on the main feed (user journey 1), 49.78\% of X's data traffic when posting a tweet (user journey 2), and 55.15\% of X's data traffic when scrolling on the TechCrunch feed (user journey 3). In absolute terms, it amounts to 117.66 MB for user journey 1 which has an average duration of 6 minutes, 6.77 MB for user journey 2, and 10.92 MB for user journey 3 that has an average duration of 6 minutes and 38 seconds. The corporate overhead is approximately $0.019156\ GB/min$ when scrolling on the main feed of X, $0.006612\ GB$ when posting a tweet, and $0.001608\ GB/min$ when scrolling on TechCrunch's feed.}

\subsubsection{User Tracking Overhead}

\rev{User tracking overhead refers to the proportion of network traffic attributable to user data collection.} It is a finer grained approach that relies on inspecting URLs and payloads for each network request. We classify the HTTP requests and responses into three categories of computational resource usage (see fig. \ref{fig:resourceproportion}). There is a fourth extra category, "other", for anything that cannot be classified with certainty. This last category is important as classifying X's network traffic is difficult and requires "reverse-engineering" the purpose of the traffic based on its content. We note that this exercise is much simpler for Mastodon as we have access to the full documentation of its API methods.\footnote{\url{https://docs.joinmastodon.org/}}

A closer inspection of X's network traffic reveals repeated requests at regular intervals to an endpoint of particular interest: \verb|/i/api/1.1/graphql/user_flow.json|. Requests to that endpoint are HTTP POST requests, which means that data is sent to X's server through that endpoint. We will call requests to that endpoint \textit{user flow requests}. User flow requests contain data with key-pair values such as \verb|action: "become_inactive"| when the user stops moving the mouse for few moments. User flow requests also contain key-pair values, like \verb|action: "impression"|, which indicates a strong relationship with user tracking and targeted advertising. The word "impression" is a common term used by user tracking and analytics tools that refers to specific content being  displayed on a user screen. Impression is a metric that means revenue in the same way "clicks" do.

\begin{figure}[h]
    \centering
    \includegraphics[width=\linewidth]{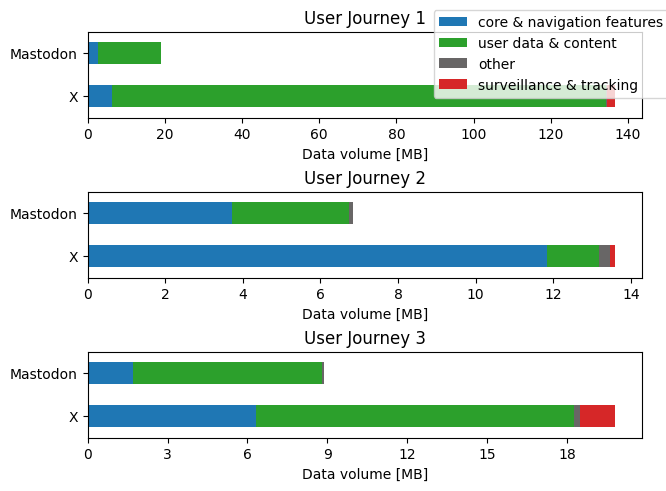}
    \caption{\rev{Bar chart showing the breakdown of network traffic for user journeys 1-3 into the different types of functionality the traffic is contributing to.}}
    \label{fig:usertrackingoverhead}
\end{figure}

Figure \ref{fig:usertrackingoverhead} shows the proportion of each category of computational resources used during the user journeys. In this case, our finer grained approach shows that surveillance and tracking account for \rev{1.66\%, 1.09\%, and 6.65\% of X's traffic for user journey 1, 2, and 3 respectively. The user tracking overhead of user journey 1 is 2.27 MB for 6 minutes, 152.48 kB for user journey 2, and 1.32 MB in a duration of 6 minutes and 38 seconds for user journey 3. Finally, we can approximate the user tracking overhead to $0.000370\ GB/min$ when scrolling on the main feed of X, $0.000145\ GB$ when posting a tweet, and $0.000193\ GB/min$ when scrolling on TechCrunch's feed.}

\subsection{Environmental Impact of Network Traffic Overheads}
\label{sec:envimpactnetworktrafficoverhead}

By using the assumption of the energy intensity coefficients and scale of X stated in section \ref{sec:envimpactquant}, the following two formulas can be used to compute yearly \cooe emission of the network traffic overhead. With $\rev{\theta_3}$ being the number of GB/min transmitted while browsing content by scrolling (user journey \rev{3}) and $\theta_2$ being the amount of GB per tweet (user journey 2):

\[(\rev{\theta_3} \times 8 \times 10^9)_{[\frac{GB}{day}]} \times 0.0591_{[\frac{kWh}{GB}]} \times 445_{[\frac{g \cooe}{kWh}]} \times 365_{[\frac{day}{year}]} \]

\[(\theta_2 \times 500 \times 10^6)_{[\frac{GB}{day}]} \times 0.0591_{[\frac{kWh}{GB}]} \times 445_{[\frac{g \cooe}{kWh}]} \times 365_{[\frac{day}{year}]} \]

\paragraph{Corporate Overhead}
\rev{We use the total corporate overhead data transferred of 0.006612 GB for a tweet (user journey 2) and 0.001608 GB/min for browsing content by scrolling (user journey 3). Computing the yearly emission for the corporate overhead for both user journeys 2 and 3 yield 414.222 t\cooe and 123.465 kt\cooe respectively. Thus, the total yearly carbon emissions of X due to corporate overhead are at least 123.879 ktCO\textsubscript{2}e.}

\rev{If we assume user journey 1 to be a representative sample of the most typical use of X and that the variability of the algorithmic feed should be accounted for by the corporate overhead, then we can assume that user journey 1 provides a suitable basis to compare X and Mastodon. From those assumptions, the results of the experiment suggest that the total yearly carbon emissions of X attributable to corporate overhead would be at least 1.471 MtCO\textsubscript{2}e.}

\paragraph{\rev{User Tracking Overhead}}
\rev{The finer grained user tracking overhead data transferred of 0.000145 GB for a tweet (user journey 2) and 0.000193 GB/min for browsing content by scrolling (user journey 3) result in the the two figures of 9.110 t\cooe and 14.877 kt\cooe for user journey 2 and 3 respectively. Adding these two figures yield a lower bound of 14.886 kt\cooe emitted yearly by X due to user tracking overhead.}

\section{Discussion}

\subsection{Obfuscation and Transparency}

Beyond the extraction of behavioral traces, CSMs following the logic of accumulation dictated by surveillance capitalism are driven by an imperative to maintain a knowledge asymmetry \cite{zuboffAgeSurveillanceCapitalism2019}. This manifests itself in the form of actively obfuscating their data extraction practices. CSMs must conceal what data is being collected, how it is being collected, and the extent of how much they have already collected.

The non-trivial nature of our measurement setup (see fig. \ref{fig:userjourneytool} and fig. \ref{fig:userjourneytoolimpl}), which required technical expertise to set up a transparent HTTPS proxy and interpret its output, illustrates the extent of underlying obfuscation, revealing only the “tip of the iceberg.” A further example is the difference between X and Mastodon in classifying HTTP request types according to the categories shown in figure \ref{fig:resourceproportion}. As referenced above, Mastodon provides documentation\footnote{\url{https://docs.joinmastodon.org/}} that enables a precise interpretation of HTTP requests. In contrast, the function of individual HTTP requests in X is largely opaque to external observers and, without insider knowledge, can only be inferred with limited confidence based on domain-specific expertise.

These types of obfuscations have consequences. When combined with complicated tasks, such as quantifying environmental impacts or social harms, the lack of transparency makes it impossible for the public to obtain detailed and accurate reports on these harms. The consequent lack of accountability facilitates under-reporting of \rev{environmental metrics and harms \cite{niranjanUSTechFirms2026}}. Combined with the political leverage afforded by surveillance capitalism to those corporations, this contributes to a catastrophic acceleration of the environmental crisis. A case in point is the recent AI boom that drives energy demands to new heights with little to no opposition from the political sphere \cite{varoquaux_hype_2025,hernandez-garcia_irresponsible_2025}. Companies like Google are admitting a resurgence of their carbon emission for the same reasons. At this point, \rev{due to the absence of transparent figures}, the public can do little more than speculate about the impacts of the heavy push for Grok with products like the Grok chatbot\footnote{\url{https://grok.com/}}, Grokipedia\footnote{\url{https://grokipedia.com/}}, and the switch to Grok-based recommendation for all X users.\footnote{\url{https://github.com/xai-org/x-algorithm}}

\subsection{Sufficient Digital Platforms}

In addition to being more transparent, FOSS ASMs, provide a useful comparison with their CSM counterparts. The absence of corporate incentives removes the need to track users for profit and substantially reduces the pressure to design addictive and engaging user experiences. ASMs are also less driven to follow the latest platform trends, such as integrated AI features or short video formats popularized by TikTok.  

Our case study focuses on the user-facing network traffic aspect of Mastodon and X; consequently, the severe effects of AI systems, such as Grok, on resource  consumption are not reflected in our results. However, X's push for a short-form, autoplaying video content can be seen in our results. In fact, we believe that this design choice is responsible for most of the observed corporate overhead in user journey 1 (see fig. \ref{fig:corporateoverhead}), \rev{This observation reinforces the interpretation that corporate overhead is a good measure of surveillance capitalism. Empirical research in business and communication studies has demonstrated a correlation between media richness and social media engagement \cite{moranMessageContentFeatures2019,caoUnderstandingConsumersSocial2021}. Because higher engagement on social media is linked to higher advertisement revenue, the introduction of increasingly rich media, such as high resolution images and short video format, is part of surveillance capitalism processes.}

Our estimation of corporate overhead's network traffic assumes that a comparable amount of time is spent by users on both platforms. However, it is also fair to assume that an ASM, such as Mastodon, that is not purposefully designed to maximize user engagement, is likely to be used less intensively than its CSM counterpart. As other scholars have noted, this makes Mastodon a suitable candidate to explore strategies of digital sustainability and sufficiency \cite{laser2022environmental}. Targeting corporate overhead with sufficiency measures, by considering FOSS ASMs as sufficient alternatives, represents a compelling approach in reducing both social and environmental harms associated with CSMs.

Another target for sufficiency measures could be the user tracking overhead. Our results show that the carbon footprint of user tracking overhead is one order of magnitude lower than that of corporate overhead. Making it a less interesting target in terms of carbon emission reduction, but may nevertheless prove to be an easier target for policy and political measures, since it does not directly challenge corporate incentives. However, such a strategy for sufficiency runs the risk of \textit{stewarding for status quo} and requires a more nuanced approach, namely one that seeks to \textit{steward for safe emergence} \cite{laurellthorslundMetacrisisComputingYou2025}. A sensibility oriented towards safe emergence would need to look beyond the carbon footprint alone \cite{knowlesClimateChangeWhat2025}. Further, understanding the environment as deeply entangled with the social and technical is essential \cite{hiltyICTSustainabilityEmerging2015}. Drawing from sustainable HCI, combining notions of digital sufficiency and our research agenda with concepts such as Social-Ecological-Technological Systems \cite{rakovaAlgorithmsSocialEcologicalTechnologicalSystems2023} or the Push and Pull framework for measuring attentional agency \cite{wojtowiczPushPullFramework2025a} offers a promising directions for future work.


\subsection{Beyond Network Traffic: X and Mastodon}

The method used in our case study, quantitatively comparing two functionally equivalent digital platforms with different business models, shows promise. We demonstrate a limited application of this approach by focusing on a single element (i.e. network traffic) of our conceptual framework (see fig. \ref{fig:framework}) and on two digital platforms (i.e. X and Mastodon). We believe that this method can be generalizable along two dimensions.

First, we can move beyond “the tip of the iceberg” by investigating additional elements of our conceptual framework. This opens further avenues for research on sufficiency within  ICT systems. For example, a study exploring the environmental impacts of corporate overhead on user devices and application UI could reuse Mastodon and X as the target of user journeys, while adopting an experimental setup similar to Puhtila et al. \cite{puhtilaEffectAnalyticsTools2024}. Another direction could explore those environmental impacts from the server-side. Within our conceptual framework, elements such as the application servers or the data storage could be evaluated by comparing a platform like Mastodon with its corporate counterpart. The fact that Mastodon is FOSS allows researchers to deploy their own instance in "lab settings" to control various parameters (e.g. the number of users or the activity of the users on the server). However, the challenges of corporate obfuscation remain; for instance, measuring the application servers' corporate overhead of a corporate digital platform would be difficult. Indeed, corporate servers are typically located within tightly controlled and highly secured  data centers with tight security.

Second, there are many other platforms like Mastodon. The Fediverse is a collection of intercommunicating ASMs based on the ActivityPub protocol.\footnote{\url{https://activitypub.rocks/}} Many ASMs on the Fediverse find their equivalent among CSMs. A non-exhaustive list of examples are the following: X and Mastodon, YouTube and PeerTube, Instagram and PixelFed, and Reddit and Lemmy. Each of these pairings offer case studies to examine the influence of corporate incentives on resource consumption. In addition, platforms such as Mastodon have "lightweight" alternatives that aim to provide functionalities but with a lower computational resource footprint. Two such alternatives are GoToSocial\footnote{\url{https://gotosocial.org/}} and Snac2\footnote{\url{https://codeberg.org/grunfink/snac2}}. Comparing these systems could provide further insights into sufficient digital platforms.

\section{\rev{Limitations}}


\subsection{\rev{Conceptual Framework}}

\rev{As the authors have no access to the backend architecture of X, the conceptual map is necessarily based on informed assumptions from the authors, who have experience with software engineering and IT infrastructure. Due to the opaque nature of surveillance capitalism processes, this map may overlook relevant elements of the system.} 

\subsection{\rev{Case Study}}

\rev{Our case study relies on several assumptions. First, the user journeys were based on basic features (e.g., scrolling, posting), rather than empirically validated patterns of usage, and may not reflect real-word user behavior. Future work could extend the existing framework and codebase for user journeys to model more complex and realistic usage patterns.}

\rev{Network traffic is an imperfect indicator of environmental impact \cite{koomeyDoesNotCompute2021}, and the assumption of a linear relationship between network traffic and carbon emissions is only appropriate for high-level approximations. Future work should better characterize the relationship between our conceptual framework components and their resource demands, as network traffic captures only one aspect of the underlying material infrastructure of surveillance capitalism.}

\rev{The numbers in section \ref{sec:envimpactnetworktrafficoverhead} are conservative as they only quantify the environmental impact of data transmission. These data are not just transmitted: they are also computed, stored, analyzed, recombined, duplicated, and sold. In each of these steps, larger data volumes translate into higher computation, storage, and network use. The environmental impact accruing from each byte of overhead is therefore higher than our lower bound estimate. Future work aiming to quantify the aforementioned steps that are unaddressed in this paper could leverage our contributed conceptual framework.}

\rev{The categorization of \textit{core \& navigation features}, \textit{user data \& content}, \textit{other}, and \textit{surveillance \& tracking} was based on heuristic criteria, such as images and videos as \textit{user data \& content} or JavaScript and HTML files as \textit{core \& navigation features}. A more structured classification of network requests would strengthen the results. Classifying requests into \textit{surveillance \& tracking} is particularly challenging, as certain HTTP headers (e.g., cookies) may serve both functional and tracking purposes. Similarly, requests that progressively stream video content may reduce bandwidth usage while also enabling fine-grained behavioral tracking (e.g., watch time). A deeper qualitative analysis of the platform traffic would improve categorization accuracy. Such an approach, however, would remain sensitive to platform updates.}

\rev{Lastly, triangulating our estimates with sustainability reports from X would strengthen confidence in the results. However, X does not publicly report its global carbon emissions. An alternative is to leverage open source intelligence and disclosures to infer operational impacts; however, such data are limited as X is now a privately held corporation. Ultimately, regulation mandating transparency in data center resource use and emissions is required.}

\section{Conclusion}

Surveillance capitalism processes have received extensive criticism for their manipulation, extraction, and pervasive surveillance tracking. In this paper, we take a first step toward quantifying their environmental impact. We introduce a conceptual framework that also functions as a research agenda for investigating the environmental impacts of surveillance capitalism. The framework provides a basis for estimating its environmental footprint through measuring corporate overhead

This case study compares network traffic between X and Mastodon. We identify surplus data volume transmitted on X: \textit{corporate overhead} and a finer grained overhead isolating traffic attributable to user tracking. Quantifying both allows us to establish two distinct lower bounds for the carbon emission associated with X's corporate practices.

We have highlighted the lack of transparency and active obfuscation of CSMs, the different approaches to digital sufficiency for digital platforms, and the potential generalizability of our contributions. More importantly, this research identifies an important knowledge gap. Addressing it will advance sustainability research while reducing the knowledge asymmetry essential for CSMs to continue operating with little to no resistance. Further research in this direction can improve accountability and help mitigate the environmental impacts of ICT as currently designed and deployed.

\section*{Acknowledgments}
Thanks to Nadia Smith for creating figure \ref{fig:resourceproportion}. This research was partially supported by NSERC through RGPIN-2025-07063, the Canada Foundation for Innovation, and the Ontario Research Fund.

\bibliographystyle{IEEEtranN}
\bibliography{refs}

\end{document}